\documentclass[manuscript]{aastex}
\usepackage{color}
\usepackage{graphicx}
\usepackage{bm}
\usepackage{amsmath}

\begin{document}
\title{Effect of cross-redistribution on the resonance scattering polarization
of O {\sc i} line at 1302 \AA\,}
\author{L.~S.~Anusha$^{1}$, K.~N.~Nagendra$^{2}$ and H. Uitenbroek$^{3}$}
\affil{$^1$Max Planck Institute for Solar System Research, Justus-von-Liebig-Weg 3,
37077 G\"ottingen, Germany}
\affil{$^2$Indian Institute of Astrophysics, Koramangala,
2nd Block, Bangalore 560 034, India}
\affil{$^3$National Solar Observatory, PO Box 62, Sunspot, NM 88349}

\email{bhasari@mps.mpg.de,knn@iiap.res.in,huitenbroek@nso.edu}

\begin{abstract}

Oxygen is the most abundant element on the Sun 
after Hydrogen and Helium. The intensity spectrum of resonance lines
of neutral Oxygen 
namely O {\sc i} (1302, 1305 and 1306 \AA\,) has been studied in the literature 
for chromospheric diagnostics.
In this paper we study the resonance scattering polarization 
in the O {\sc i} line at 1302 \AA\, using two-dimensional radiative transfer in 
a composite atmosphere constructed using a two-dimensional 
magneto-hydrodynamical snapshot in the photosphere and columns of the one-dimensional FALC atmosphere in the 
chromosphere. The methods developed by us recently in a series of 
papers to solve multi-dimensional polarized radiative transfer have been 
incorporated in our new code POLY2D which we use for our analysis. 
We find that multi-dimensional radiative transfer including 
XRD effects is important in  reproducing the amplitude and shape of
scattering polarization signals of the O {\sc i} line at 1302 \AA\,. 
\end{abstract}

\section{Introduction}
\label{intro}
Frequency cross redistribution (XRD) has been introduced in the literature to
take into account the effects of partial frequency redistribution (PRD) in multi-level 
atomic systems. The O {\sc i} triplet has been studied in great 
detail by \citet[][]{carlssonjudge1993} using multi-level radiative transfer.
\citet[][]{elizahu2002} have shown the importance of XRD in modeling the 
intensity spectrum of these lines. However, the effects of XRD on scattering 
polarization have not been quantified so far. In this paper we 
investigate such effects in the O {\sc i} resonance line at 1302 \AA.

The XRD theory is developed in the classic papers 
by \citet[][]{hubenyetal83a,hubenyetal83b}. 
Recently, \citet[][]{sametal13} showed a heuristic
way of including XRD to study the scattering polarization in multi-level
atoms. Rigorous QED theory that can include XRD 
effects on the polarization is under 
development by several authors in the field \citep[see e.g.,][]{bom2003}. 
However they are not yet published with complete details. 

Given the theoretical difficulties to solve the full 
polarized multi-level transfer equation rigorously, we 
decompose the problem as follows. Small degrees of scattering 
polarization in O {\sc i} lines allow us to assume that a decoupling
of the underlying transfer equation into an unpolarized multilevel atom 
transfer equation with XRD, and a polarized two-level atom transfer equation 
with ordinary PRD \citep[i.e., no XRD, following][]{bom97a,bom97b}, is 
sufficiently good for obtaining an estimate of the effects of 
XRD on the emergent, fractional linear polarization. 

We present a comparison of intensity and linear polarization profiles 
computed by assuming either ordinary PRD or XRD in the multi-level calculation,
and show the significant differences in the shapes and amplitudes of 
the fractional linear polarization signals.

\section{Theoretical background}
\subsection{The Stokes parameters}
\begin{figure*}
\centering
\includegraphics[scale=0.5]{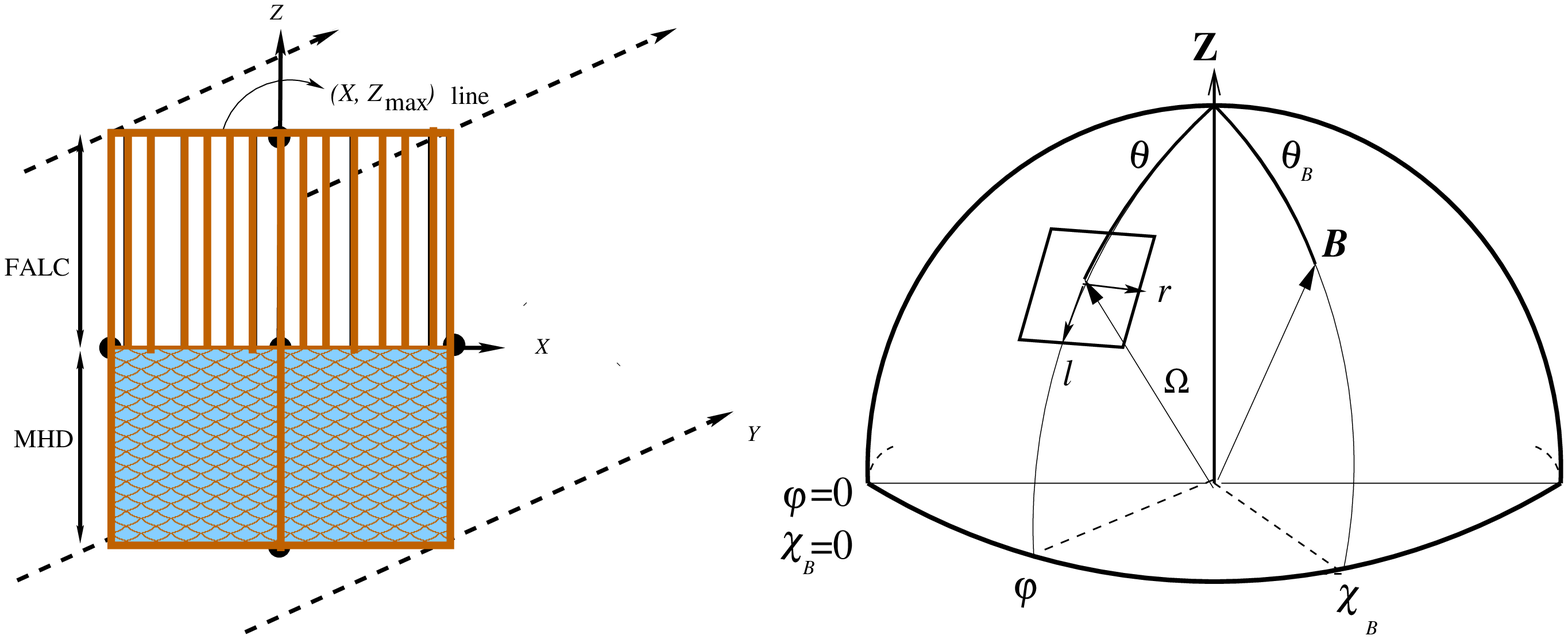}
\caption{Left: Geometry of the 2D radiative transfer problem. 
The cartoon mimics the spatially inhomogeneous MHD atmosphere in
the photosphere and horizontally homogeneous columns of 
FALC atmosphere in the chromosphere. The line $(X,Z_{\rm max})$
represents the surface where the radiation emerges from the 2D atmosphere. 
Right: The atmospheric reference frame.
The angle pair $(\theta,\varphi)$ defines
the outgoing ray direction. The magnetic field is
characterized by $\bm{B}=(\Gamma,\theta_B,\chi_B)$, where $\Gamma$ is
the Hanle efficiency parameter and ($\theta_B,\chi_B$)
defines the field direction. $\Theta$ is the scattering angle.}
\label{fig-001}
\end{figure*}

We consider an ellipse
described by vibrations of an electric field vector corresponding to
an elliptically polarized beam of light \citep[see][]{chandra60}.
Consider a plane transverse to the propagation direction of this beam of light
in which we can decompose the specific intensity into components $I_l$ and $I_r$ along two mutually
perpendicular directions $l$ and $r$ (see Figure~\ref{fig-001}, right panel). 
Then we define
\begin{eqnarray}
I=I_l+I_r,  \nonumber \\
Q=I_l-I_r, \nonumber \\
U=(I_l-I_r) \tan 2 \chi,
\label{stokesIQU}
\end{eqnarray}
where $\chi$ is the angle between the semi-major axis of the ellipse and
the direction $l$. Positive value of $Q$ is
defined to be in a direction parallel to $l$ and negative $Q$
to be in a direction parallel to $r$. The vector 
$\bm{r}=({\rm x},{\rm y}, {\rm z})$
denotes the position vector of the ray described by the direction cosines 
$\bm{\Omega}=(\eta,\gamma,\mu)=
(\sin\theta\,\cos \varphi\,,\sin\theta\,\sin \varphi\,,\cos \theta)$ 
with respect to the atmospheric normal (the $Z$-axis in Figure~\ref{fig-001}, left panel). 
Here $\theta$ and
$\varphi$ represent the polar and azimuthal angles of the
ray (see Figure~\ref{fig-001}, right panel).
 
\subsection{The cross-redistribution (XRD)}
In this section we briefly describe the theoretical background
of the XRD mechanism. For more details we refer the reader 
to \citet[][]{hubenyetal83a,hubenyetal83b} 
\citep[see also][and the references cited therein]{hu01}. 

As described in \citet[][]{hu01}, for a ray travelling in direction $\bm{\Omega}$,
at wavelength $\lambda$, the emission profile coefficient denoted as $\psi$, in the case of 
a general XRD theory depends on the radiation field $I(\lambda,\bm{\Omega})$ in the line and 
its subordinate lines. It is given by 
\begin{eqnarray}
\left[\psi_{ij}(\lambda,\bm{\Omega})\right]_{\rm XRD} = 
&&\phi_{ij}(\lambda,\bm{\Omega})\Big\{1+\frac{\sum_{k<j}n_k B_{kj}}{n_j P_j} \nonumber \\
&& \times \oint \frac{d\bm{\Omega}'}{4\pi} \int d\lambda' I(\lambda',\bm{\Omega}') 
 \left[\frac{R_{kji}(\lambda,\bm{\Omega},\lambda',\bm{\Omega}')}{\phi_{ij}(\lambda,\bm{\Omega})}
-\phi_{kj}(\lambda',\bm{\Omega}')\right]
\Big\}.
\label{psi-xrd}
\end{eqnarray}
In the case of ordinary PRD the summation over all subordinate lines in the 
emission profile coefficient reduces to just one term 
with $k=i$.

Here $n_i, n_j, n_k$ denote populations of $i,j,k$-th levels respectively; $R_{kji}$ are
the cross-redistribution functions involving levels denoted by the subscripts $k$, $j$ and 
$i$; $\phi_{ij}$ denotes the Voigt profile function for the line corresponding to transition 
between levels $j$ and $i$; $B_{kj}$ or $B_{ji}$ being the Einstein-B coefficients for
the transitions described by the subscripts $j$ and $i$; and $P_j$ denotes the sum of 
collisional and radiative rates, namely $P_j=\sum_{k \ne j} C_{jk}+R_{jk}$. 

We note here that although general expressions given above 
can handle angle-dependent XRD/PRD problems, we restrict our attention
in this paper to the angle-averaged redistribution functions 
\citep[see][]{hum62}.

\section{Radiative transfer in the O {\sc i} triplet}
\label{method}

The method of solution used in this paper to compute the fractional linear polarization 
in the O {\sc i} resonance line is the same as that explained  
in \citet[][and references cited therein]{anuknn13}. For completeness we summarize
the method briefly in this section.

\begin{figure*}
\centering
\includegraphics[scale=0.5]{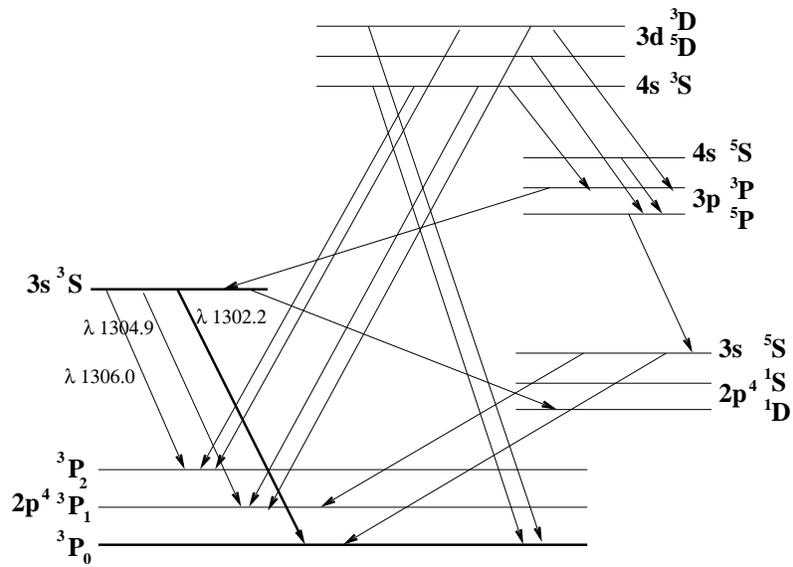}
\caption{Grotrian diagram of the O {\sc i} triplet showing all the eighteen line transitions 
that are taken into account in the unpolarized radiative transfer calculations
in the RH-code. The 1300 triplet line transitions are labelled with the line center
wavelengths. The energy levels of the resonance line transition at 1302.2 \AA\, are 
marked in bold lines the diagram.}
\label{fig-002}
\end{figure*}

In the first step we solve 
unpolarized transfer with the multi-level code of Uitenbroek
\citep[see][hereafter called the RH code]{hu01}
with the O {\sc i} model atom and a model atmosphere which is a
combination of 
a two-dimensional (2D) 
snapshot of a three-dimensional (3D) magneto-hydrodynamical (MHD) 
atmosphere in the photosphere \citep[see][]{nordstein91}, 
and columns of the one-dimensional FALC atmosphere in the chromosphere \citep[see][]{fontenla93}. 
For the solution of unpolarized multi-level transfer equation 
with XRD the model atom used is the same as that used in \citet[][]{elizahu2002}. 
It has 14 atomic energy levels with 18 line 
transitions and 13 continuum transitions in O {\sc i}. 
A Grotrian diagram of the O {\sc i} triplet and the other lines taken into account
when solving with the RH-code is shown in Figure~\ref{fig-002}. 
The continuum transitions are not shown. The main lines of the triplet are 
results of the transitions 
($3s\,\,^3S \rightarrow {2p^{4}}\,\, ^{3}P_0$ :\,1302.2 \AA\,),
($3s\,\,^3S \rightarrow {2p^{4}}\,\,^{3}P_1$ :\,1304.9 \AA\,), and 
($3s\,\,^3S \rightarrow {2p^{4}}\,\, ^{3}P_2$ :\,1306 \AA\,). 
The XRD theory of \citet[][]{hubenyetal83a,hubenyetal83b} is 
incorporated in the unpolarized transfer equation in the RH code. 
The RH-code computes radiative rates, collision rates, center-to-limb variation of
the intensity spectrum $I(\lambda,\bm{\Omega})$ etc. in the O {\sc i} triplet. 

In the second step we only consider the resonance line (marked in bold 
in Figure~\ref{fig-002}), namely the line at 1302\,\AA\,, which is a result of the
resonance transition $3s\,\,^3S \rightarrow {2p^{4}}\,\, ^{3}P_0$.
All the output parameters from the RH-code corresponding to this line 
are kept fixed in this step. 
For the fractional linear polarization computation in the second step, 
we adapt the irreducible spherical tensor notation \citep[see e.g.,][]{ll04,hf07}.
We use $I(\lambda,\bm{\Omega})$ from the RH-code
for both ordinary PRD and XRD cases as the initial unpolarized solution namely, the 6-component
irreducible Stokes vector $\bm{\mathcal{I}}=(I^0_0,0,0,0,0,0)^T$ \citep[see e.g.,][]{anuknn13}.
We then calculate the 6-component mean intensity vector in the two-level atom 
approximation with an unpolarized ground level given by 
\begin{eqnarray}
&&\bm{\mathcal{J}}(\lambda)=\frac{1}{\phi(\lambda)} 
\int_{-\infty}^{+\infty} d\lambda' \oint\frac{d\bm{\Omega}'} {4 \pi} 
\hat{W}\Big\{\hat{M}_{\rm II}(\bm{B},\lambda,\lambda')
r_{\rm II}(\lambda,\lambda') \nonumber \\
\!\!\!\!\!\!&&+\hat{M}_{\rm III}(\bm{B},\lambda,\lambda')
r_{\rm III}(\lambda, \lambda') \Big\}
\hat{\Psi}(\bm{\Omega}') \bm{\mathcal{I}}(\lambda',\bm{\Omega}').
\label{jbar}
\end{eqnarray}
Here $\bm{B}$ represents an oriented magnetic field vector (see 
Figure~\ref{fig-001}, right panel).
The matrix $\hat{\Psi}$ represents the reduced phase matrix for
Rayleigh scattering. Its elements are listed in Appendix D
of \citet{anuknn11b}. The elements of the
matrices $\hat{M}_{\rm II,III}(\bm{B},\lambda,\lambda')$ for the Hanle
effect are derived in \citet{bom97a,bom97b}. 
The functions $r_{\rm II}$ and $r_{\rm III}$ are the
angle-averaged PRD functions of \citet[][]{hum62}, keeping
the same notation as \citet[][]{anuknn13}.
$\hat{W}$ is a diagonal matrix written as
\begin{equation}
\hat{W}=\textrm{diag}\{W_0,W_2,W_2,W_2,W_2,W_2\},
\label{w}
\end{equation}
with $W_0=1$. The angular momentum quantum number of the line under 
consideration determines the value of $W_2$ \citep[see e.g.,][]{ll04}. 
For the O {\sc i} line at 1302 \AA\,, the factor $W_2=0.01$.
Here $\phi(\lambda)$ represents the Voigt profile function for the resonance line
at 1302 \AA\,.

Computing the mean intensity vector is an important step in which linear 
polarization is generated from the unpolarized intensity vector. This step
is the link between the multi-level atom unpolarized radiative transfer code (RH-code) and the
two-level atom polarized radiative transfer code that computes linear polarization at 1302 \AA\,.
In this second step the polarization is generated iteratively similar to a perturbation process
and we finally obtain converged fractional linear polarization. 
For more details on the computation of linear polarization we refer the reader to
\citet[][and the references cited therein]{anuknn13}.
For future references we name the code used in the second step as POLY2D, 
which is a 2D extension of the so called POLY code developed by \citet[][]{flurietal2003}. 

We note here that in this paper we use the word `XRD' when we use 
cross redistribution theory in the RH-code
to compute linear polarization using two-level atom PRD theory, and 
use the word ordinary `PRD'
when we use ordinary PRD theory in the RH code to
compute linear polarization using two-level atom PRD theory.

\section{Results and Discussions}
In this section we present the intensity profiles of the O {\sc i} triplet 
and linear polarization profiles of the O {\sc i} line at 1302 \AA\, computed 
using the method described in Section~\ref{method}. Our emphasis is on the fractional 
linear polarization spectrum of the O {\sc i} line at 1302 \AA\,. We also present an estimate
of the errors in the resonance scattering fractional polarization, caused
by not using XRD. We note here that the degree of the polarization in the 
O {\sc i} resonance line is extremely small (because of the small $W_2$ factor), yet 
they are useful to theoretically predict the effects of XRD on the fractional
linear polarization. All the results presented here correspond to a non-magnetic case
($\bm{B}=0$).

\subsection{Intensity profiles}
\begin{figure*}
\centering
\includegraphics[scale=0.35]{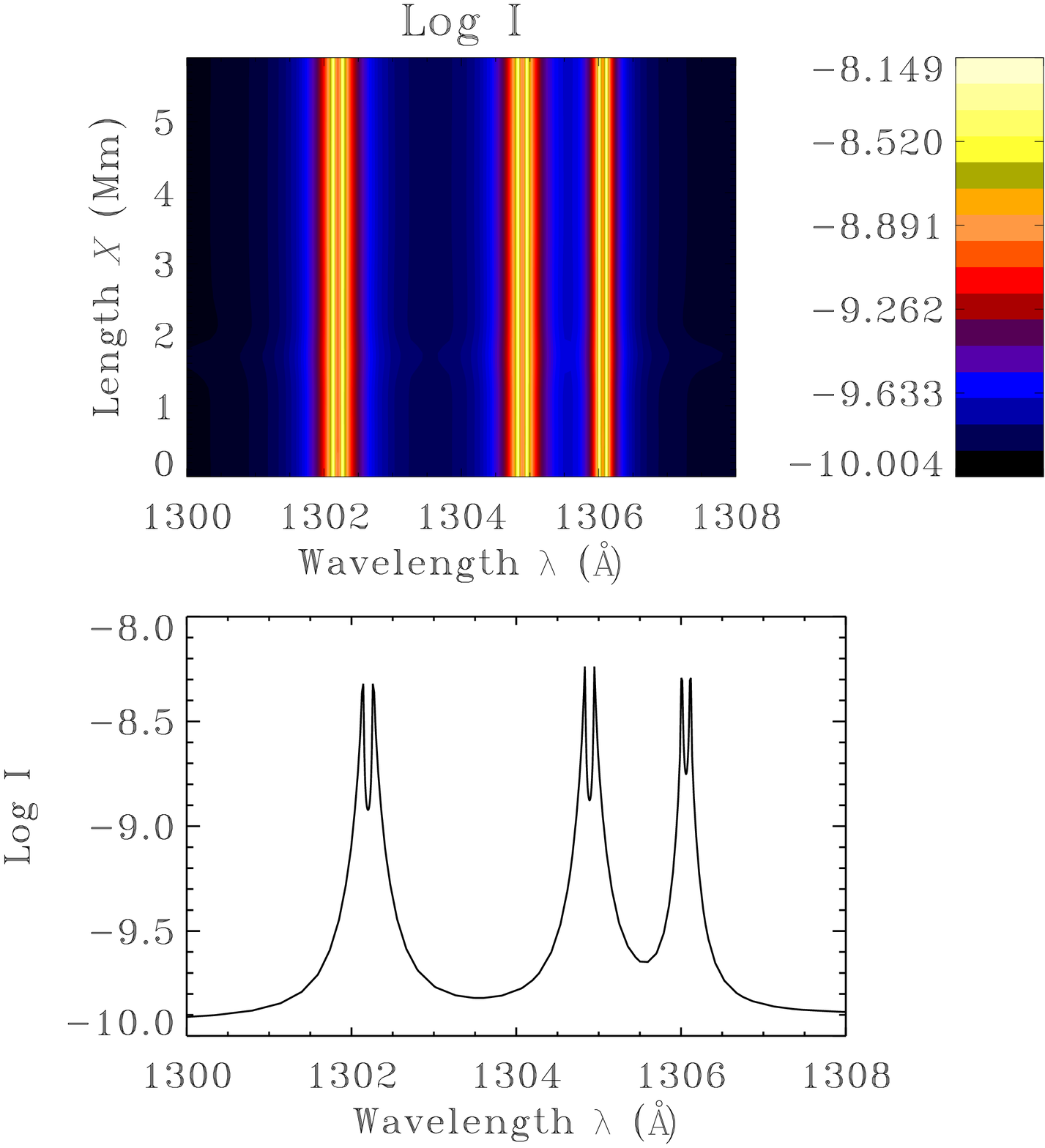}
\includegraphics[scale=0.35]{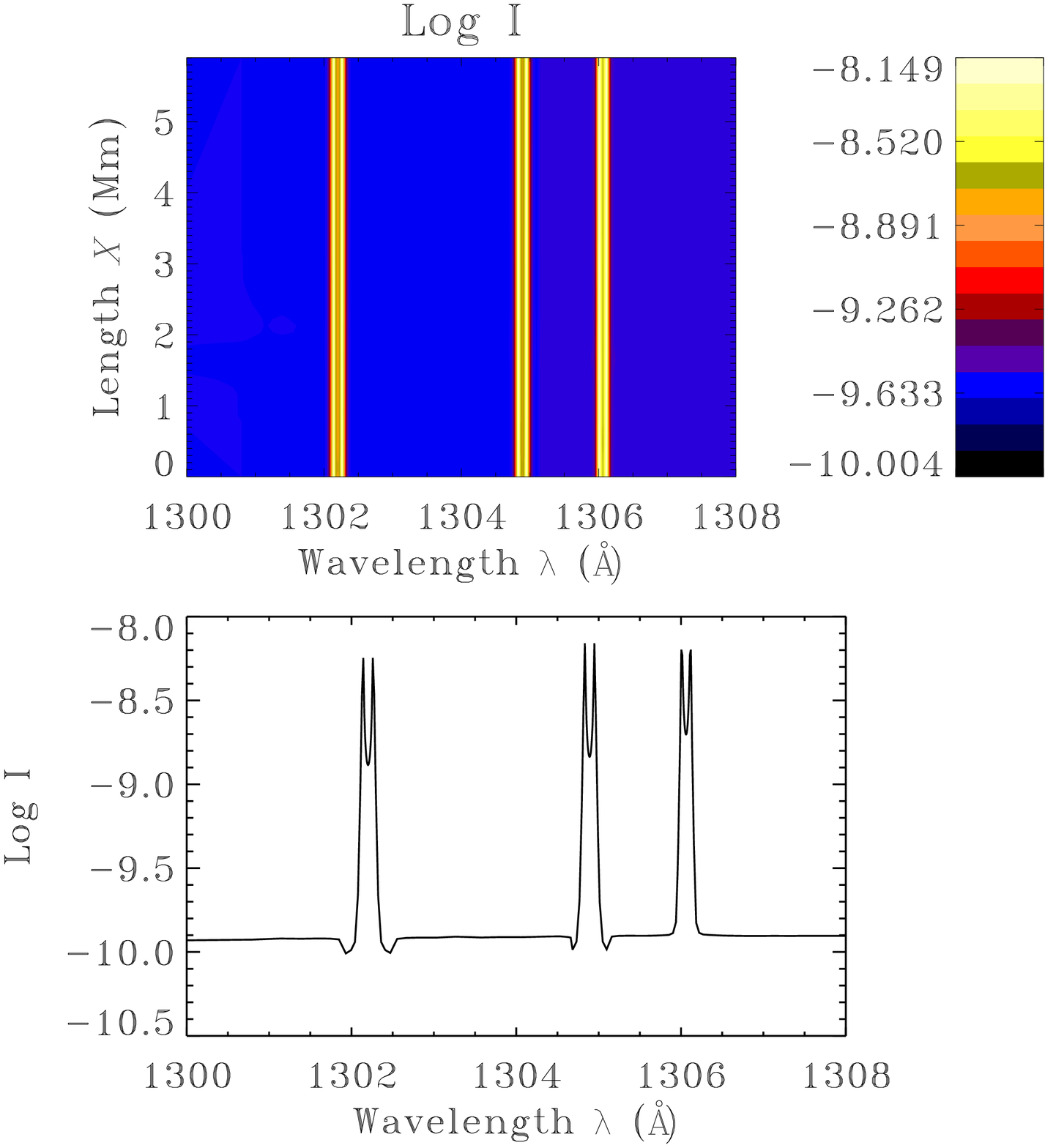}
\caption{Top: The 2D spatial variation of the emergent, intensity profiles of the 
O {\sc i} triplet line in log scale. Left panels are computed using
ordinary PRD theory and those on the right panels are computed using
XRD theory in the unpolarized multi-level radiative transfer.
Bottom: Emergent, spatially averaged intensity profiles
of the triplet in log scale.}
\label{fig-0034}
\end{figure*}

In Figure~\ref{fig-0034} we show the emergent intensity profiles of the O {\sc i} triplet in log scale
computed from the 2D version of the RH-code for ordinary PRD and the XRD cases.
We show both, namely the spatial variation of the emergent intensity profiles and 
their spatial averages. 
The color coding of the profiles in the top panels is shown
in the corresponding color bars in each panel.
The spatial homogeneity of the intensity profiles is caused by
the fact that the formation height of the line core is in a region of the atmosphere where 
it is represented by the horizontally homogeneous FALC atmosphere. 
A comparison of the intensity profiles computed using the ordinary PRD and the XRD cases 
show similar conclusions as presented in \citet[][]{elizahu2002}. 
The use of ordinary PRD instead of XRD leads to a broadening of
the intensity profiles in the near-wings of the line.
The reason is that, if we do not take XRD into account, the fraction of emissions
that occur by excitation in two other lines of the triplet is counted as 
complete frequency redistribution (CRD) in the line under consideration, and clearly 
CRD leads to a broadening in the lines. Therefore assuming 
ordinary PRD decreases the coherency fraction of the lines. 

\subsection{Linear polarization profiles}
\begin{figure*}
\centering
\includegraphics[scale=0.37]{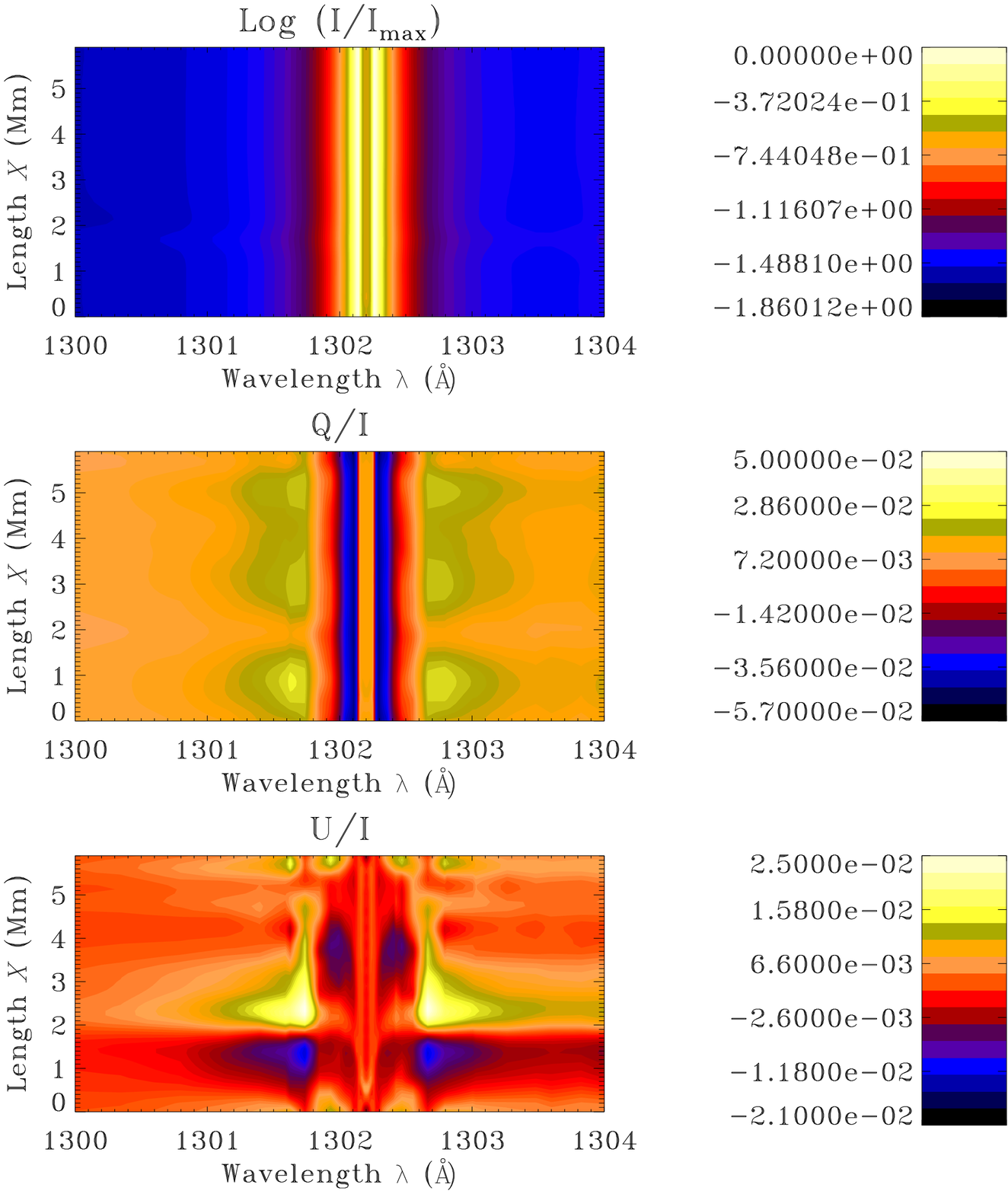}
\hfill
\includegraphics[scale=0.37]{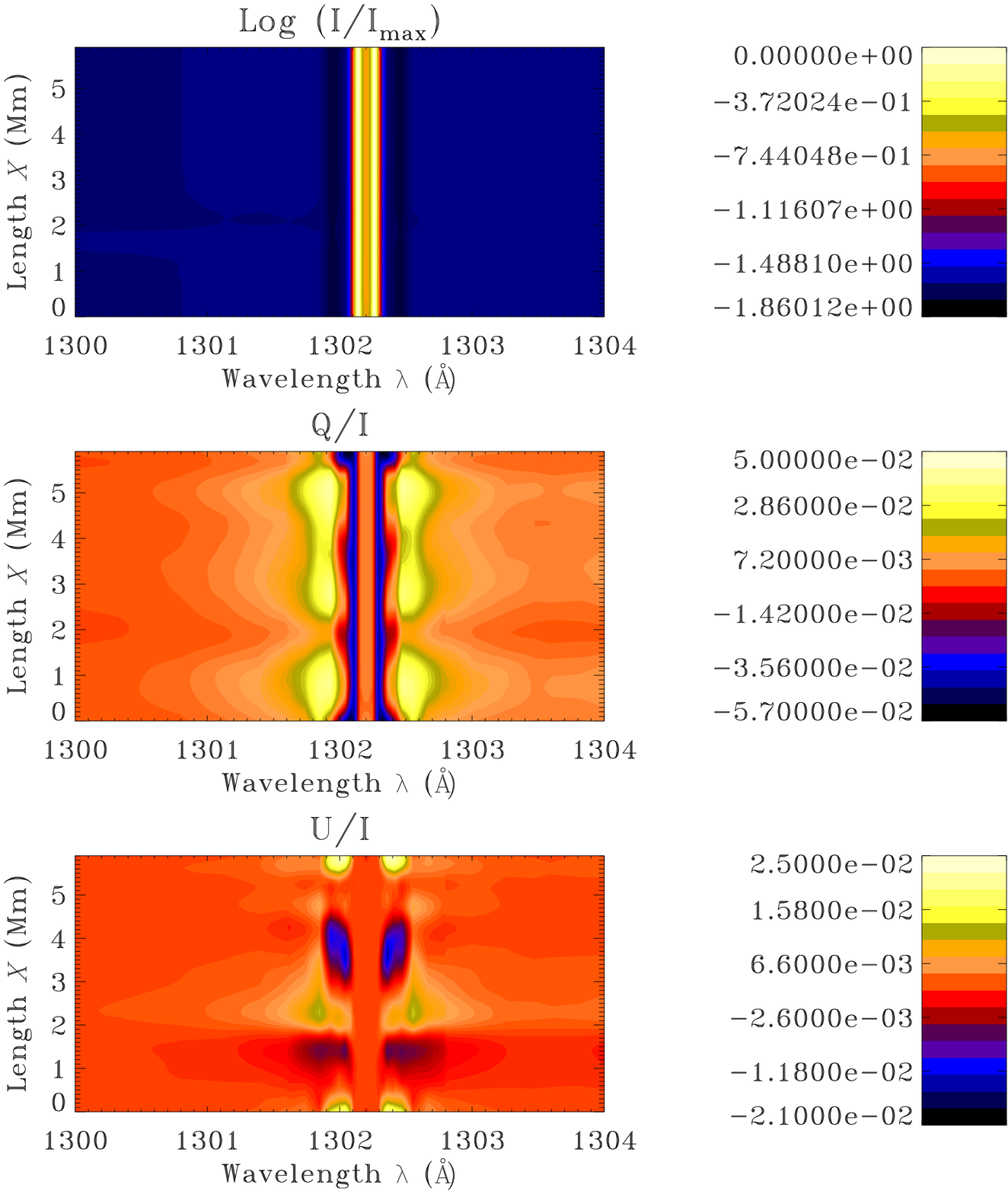}
\caption{Spatial distribution of $(I/I_{\rm max},Q/I,U/I)$ for the first
direction in Carlsson A4 angular grid \citet[][]{car63},
namely, $(\mu,\varphi)=(0.3,160^{\circ})$
on the top of the 2D atmosphere. Left panels are computed using 
ordinary PRD theory and those on the right panels are computed using 
XRD theory in the unpolarized multi-level radiative transfer.}
\label{fig-01}
\end{figure*}

In Figure~\ref{fig-01} we present a spatial distribution of 
$(I/I_{\rm max}, Q/I, U/I)$ profiles at the top of the 2D atmosphere
for both the ordinary PRD and XRD cases. The color coding of the profiles is shown
in the corresponding color bars in each panel. 
Clearly, differences between the ordinary PRD and XRD profiles are significant
in both the line core and the near-wings of the line. Unlike the intensity profiles,
the fractional linear polarization profiles are sensitive to the
structuring in the atmosphere. The spatial variation due to
inhomogeneities of the MHD atmosphere in the lower layers shows itself in 
the near-wings of the line which are formed in these layers 
\citep[see][]{anuknn13}. The spatial distribution of linear polarization
shows significant differences between the ordinary PRD and XRD cases.

\begin{figure*}
\centering
\includegraphics[scale=0.8]{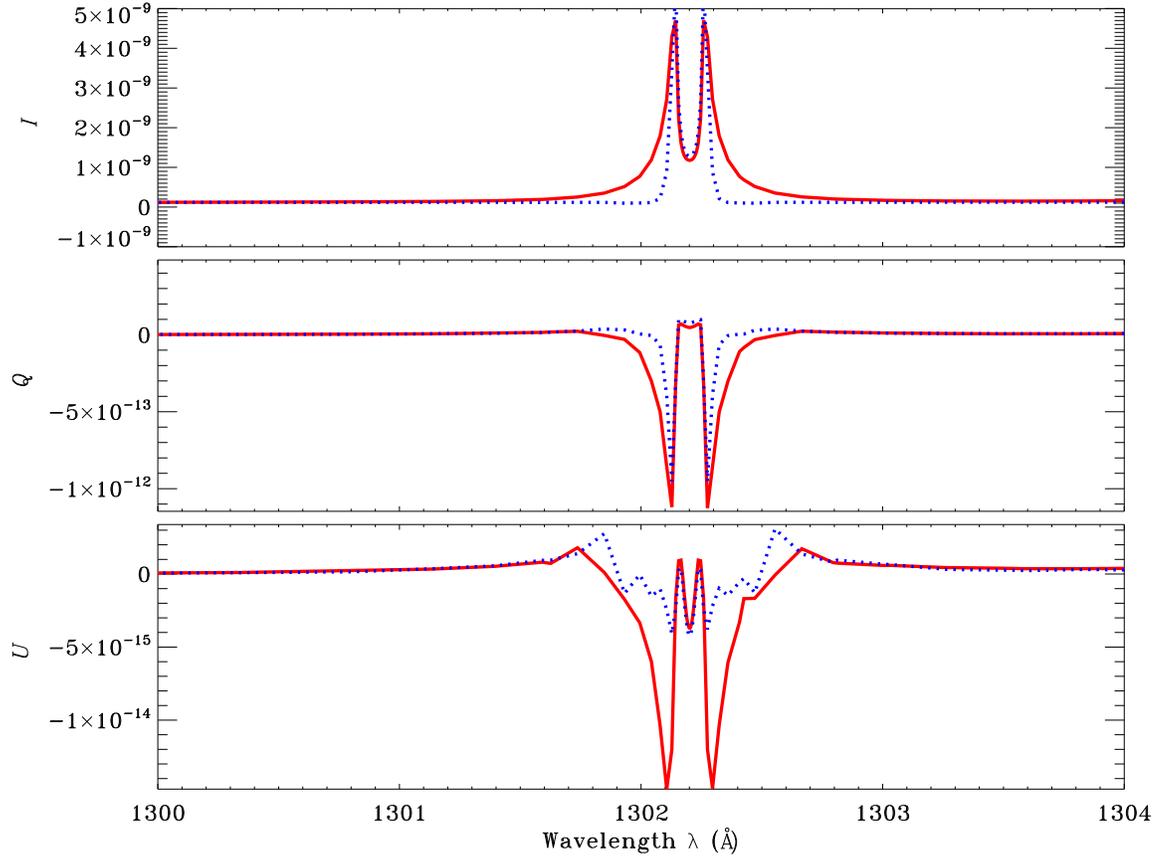}
\caption{
Spatially averaged $(I,Q,U)$ profiles for ordinary PRD (solid red curves)
and XRD (dotted blue curves) cases for $(\mu,\varphi)=(0.3,160^{\circ})$.
}
\label{fig-02a}
\end{figure*}

In Figure~\ref{fig-02} we show the spatially averaged
fractional polarization profiles $(I/I_{\rm max},Q/I,U/I)$ profiles
where each of the Stokes parameters are spatially averaged before considering
their ratios. These profile shapes are very much similar to the
spatially averaged $(I,Q,U)$ profiles for the XRD and the
ordinary PRD cases as shown in Figure~\ref{fig-02a}. However in both these cases,
the spatial averaging smears out large differences between XRD and the ordinary PRD
fractional polarization profiles, which are clearly seen in Figure~\ref{fig-01}. 
This suggests that the differences between XRD and the ordinary PRD fractional 
polarization profiles become more significant in spatially resolved structures.

The 2D radiative transfer allows us to show the effects of XRD also
in the resonance scattering values of $U/I$ - which is generated due to 
geometrical symmetry breaking in the medium. 

In the case of scattering polarization, the quantities of interest have always been, 
the fractional polarization profiles. This is because of the following reasons.
From the observational point of view, even in the most sophisticated polarimeter that 
measures scattering polarization to a very high accuracy, namely ZIMPOL 
\citep[see e.g.,][]{achimetal04}, certain noise (seeing noise and gain table noise namely 
flat-field noise due to pixel-to-pixel sensitivity variations) in 
the observed data can be removed only by 
measuring the fractional polarization values $Q/I$ and $U/I$ at each wavelength point.
The small values of $I$ in a given spectral region further makes it difficult to measure the 
polarization profiles $Q$ and $U$, because the latter are 
obtained through a decomposition of $I$ in two mutually perpendicular
directions (see Equation~\ref{stokesIQU} and Figure~\ref{fig-02a}). 
From theoretical point of view, physically more meaningful are the fractional
polarization profiles expressed in percents, in place of the $Q$ and
$U$ profiles which are expressed in some arbitrary units.

For all these reasons, we conclude that the differences between 
XRD and ordinary PRD fractional polarization profiles need to be considered when modeling 
the observed fractional polarization signals.

\subsection{Error estimation}
\begin{figure*}
\centering
\includegraphics[scale=0.8]{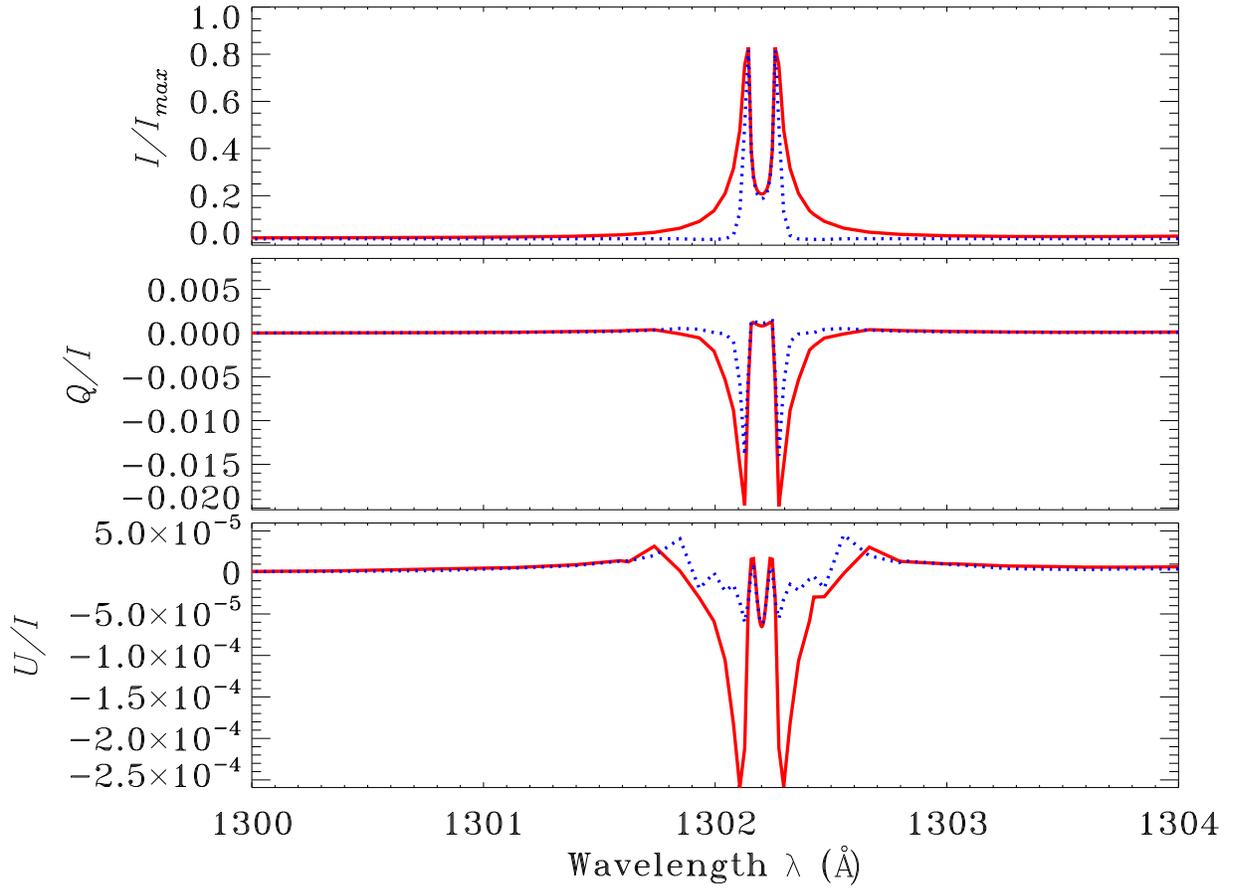}
\caption{
$(I/I_{\rm max},Q/I,U/I)$ profiles with each of the Stokes parameters spatially
averaged, for ordinary PRD (solid red curves)
and XRD (dotted blue curves) cases for $(\mu,\varphi)=(0.3,160^{\circ})$.
}
\label{fig-02}
\end{figure*}

As we discussed in the previous section, the XRD effects apparently 
become more visible in spatially resolved fractional polarization 
profiles (as seen in Figure~\ref{fig-01}), than in the spatially averaged
case (as in Figures~\ref{fig-02} and \ref{fig-02a}) as the latter smears 
out the differences. Therefore, to quantify the differences between ordinary 
PRD and XRD profiles in the spatially resolved case, 
we computed percent absolute error defined as
\begin{equation}
E=(|\,\,|X_{\textrm{XRD}}| - |X_{\textrm PRD}|\,\,|) \times 100,
\label{abs-err}
\end{equation}

where $X=(I/I_{\rm max}, Q/I, U/I)$. We do not use the relative error because
in some parts of the spectrum, values of the linear polarization
approach zero in which case, errors become undefined.

\begin{figure*}
\centering
\includegraphics[scale=0.5]{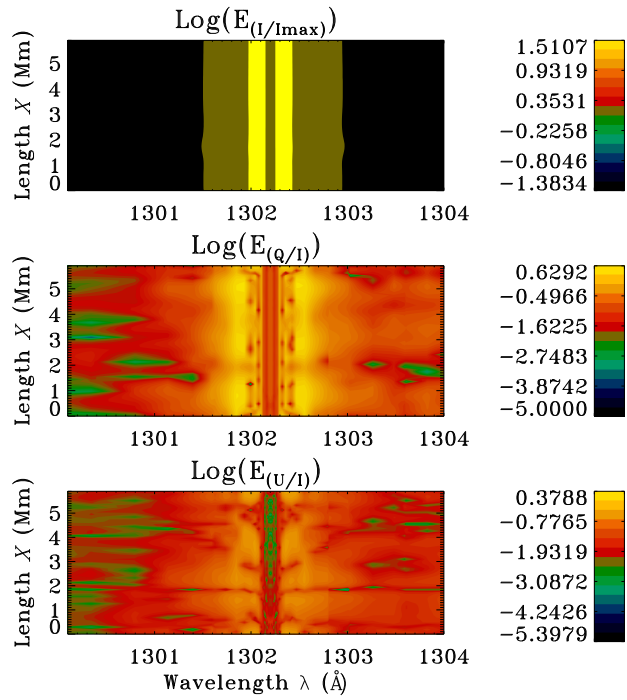}
\caption{Percent absolute errors in $(I/I_{\rm max},Q/I,U/I)$ profiles,
expressed in log scale, computed using 
Equation~(\ref{abs-err}) for $(\mu,\varphi)=(0.3,160^{\circ})$.}
\label{fig-03}
\end{figure*}

In Figure~\ref{fig-03} we show percent absolute error in $(I/I_{\rm max}, Q/I, U/I)$,
(in log scale), caused by using ordinary PRD theory instead of XRD theory. Due to 
steep gradient in the
intensity profiles (see Figure~\ref{fig-02}), the errors can be as large as
28 \% in the line wings ($\sim$1.44 in log scale) of intensity profiles. In $Q/I$ profiles, the maximum
absolute error is 4 \% ($\sim$ 0.62 in log scale) and in $U/I$, it is 2 \% 
($\sim$ 0.37 in log scale). These results clearly
show that XRD can cause significant differences in intensity as well as linear
polarization profiles, and needs to be used whenever possible
and applicable.   

On the one hand, the spatial distribution of the scattering polarization
shows that it is sensitive to the structuring of the atmosphere. On the other hand
significant differences between the spatial distribution of the polarization for the 
ordinary PRD and XRD cases show that whenever XRD is relevant, assumption of ordinary PRD
may lead to significant errors. Therefore, the differences between ordinary PRD and XRD profiles is the
outcome of both (1) multi-D transfer, and (2) the type of scattering
theory used. In this way the XRD effects in combination with multi-D transfer 
are important to reproduce the amplitude and shapes of the observed scattering
polarization profiles. 

\section{Conclusions}
In this paper we have attempted to quantify for the first time, the effects of XRD
on the fractional linear polarization using the example of 
the resonance line of O {\sc i} triplet at 1302 \AA\,. We use the anisotropy 
computed from the RH-code for both ordinary PRD and XRD
theories, to compute the linear polarization based on the 
two-level atom PRD theory of \citet[][]{bom97a,bom97b}.
We present also an estimate of 
the percent absolute error when using the approximation of ordinary 
PRD theory instead of the more realistic XRD theory. This can be as large as 28 \% for the intensity profiles, due 
to the steep gradients present in these profiles. In $Q/I$, the maximum error is 4 \% 
and that in $U/I$ it is 2 \%. 
We show a comparison of the spatial distribution of the fractional scattering polarization and the 
spatially averaged profiles for ordinary PRD and XRD cases. We conclude that 
the fractional linear polarization signals are sensitive to the structuring of the atmosphere and
that the XRD effects become more significant in spatially resolved cases than in the spatially averaged
case as the latter smears out the large differences. 
Therefore, to reproduce the amplitude and shape of observed scattering polarization signals of 
the O {\sc i} line at 1302 \AA\,, a multi-dimensional radiative transfer including 
XRD theory is important. 

\acknowledgments
\noindent
L.S.A. would like to thank the Alexander von Humboldt foundation for the fellowship that 
supported this project at MPS, G\"ottingen, Germany. We would like to thank 
Prof. J. O. Stenflo for useful suggestions and Dr. Michiel van noort 
for useful discussions.



\end{document}